**High dynamic range, heterogeneous, terahertz quantum cascade lasers featuring thermally-tunable frequency comb operation over a broad current range**

**Katia Garrasi,[1] Francesco P. Mezzapesa,[1] Luca Salemi,[1] Lianhe Li,[2] Luigi Consolino,[3] Saverio Bartalini,[3] Paolo De Natale,[3] A. Giles Davies,[2] Edmund H. Linfield,[2] and Miriam S. Vitiello[1]**

[1]*NEST, CNR - Istituto Nanoscienze and Scuola Normale Superiore, Piazza San Silvestro 12, 56127, Pisa, Italy*
[2]*School of Electronic and Electrical Engineering, University of Leeds, Leeds LS2 9JT, UK*
[3] *CNR-Istituto Nazionale di Ottica and LENS (European Laboratory for Non-linear Spectroscopy), Via Carrara 1, 50019 Sesto Fiorentino (FI), Italy*

**Abstract.** We report on the engineering of broadband quantum cascade lasers (QCLs) emitting at Terahertz (THz) frequencies, which exploit a heterogeneous active region scheme and have a current density dynamic range ($J_{dr}$) of 3.2, significantly larger than the state of the art, over a 1.3 THz bandwidth. We demonstrate that the devised broadband lasers operate as THz optical frequency comb synthesizers in continuous-wave, with a maximum optical output power of 4 mW (0.73mW in the comb regime). Measurement of the intermode beatnote map reveals a clear dispersion-compensated frequency comb regime extending over a continuous 106 mA current range (current density dynamic range of 1.24), significantly larger than the state of the art reported under similar geometries, with a corresponding emission bandwidth of ≈ 1.05 THz and a stable and narrow (4.15 KHz) beatnote detected with a signal-to-noise ratio of 34 dB. Analysis of the electrical and thermal beatnote tuning reveals a current-tuning coefficient ranging between 5 MHz/mA and 2.1 MHz/mA and a temperature-tuning coefficient of –4 MHz/K. The ability to tune the THz QCL combs over their full dynamic range by temperature and current paves the way for their use as a powerful spectroscopy tool that can provide broad frequency coverage combined with high precision spectral accuracy.

The terahertz (THz) region of the electromagnetic spectrum, conventionally covering the frequency range from 0.1 THz to 10 THz, plays an important role for applications [1]. Specifically, sensing, high-resolution spectroscopy [2-4], and metrology [5] have an enormous potential in this spectral region since many chemical compounds and simple molecules possess rotational and vibrational resonances at THz frequencies [1-3]. However, rotational transitions have natural linewidths of a few Hz and Doppler limited line-widths of a few kHz, and so many applications require narrow linewidth sources that have a precisely controlled frequency.

Quantum cascade lasers (QCLs) [6,7] are the most versatile source of THz frequency radiation, combining compactness, a broad optical gain bandwidth [8-10], continuous wave (CW) operation, high output power [11], and an inherently high spectral purity [12,13] with intrinsic linewidths as low as 100 Hz [12].

In recent years, it has been demonstrated that broadband THz QCLs can also operate as optical frequency comb synthesizers [14,9], without requiring any additional mode-locking mechanism [15]. For this reason, THz QCLs have generated interest for a wide range of possible metrological applications, such as frequency synthesis and molecular sensing or for dual-comb spectroscopy, where multi-heterodyne detection [4] is performed, allowing both high-resolution and high sensitivity to be obtained.

An optical frequency comb comprises of a set of perfectly equally spaced modes, with the associated phases locked in a well-defined relationship to each other. Theoretical [15,16] and experimental investigations [14,9,17] have confirmed that QCLs behave like frequency combs owing to intracavity third-order non-linear optical processes, stemming from a very large nonlinear susceptibility $\chi^{(3)}$ in the active material. Specifically, the resonant third-order non-linearity inherently induces self phase-locking by the four-wave–mixing (FWM) process. FWM tends to homogenize the mode spacing and consequently acts as the main mode proliferation and comb generation mechanism in a free running QCL [9, 15-17].

However, in a free-running multimode QCL, the modes are not uniformly spaced owing to chromatic dispersion. As result of a frequency-dependent refractive index, the laser free spectral range is index dependent, producing an unevenly spaced spectrum. The competition between FWM and group velocity dispersion (GVD) in the laser cavity therefore plays a major role in determining the QCL comb-degradation mechanism [14].

The total GVD in a QCL comprises three main components: the dispersion from the "Reststrahlen band" of the host semiconductor material $GVD_{mat}$; the modal dispersion of the waveguide $GVD_{mod}$; and, the dispersion associated with the intersubband transitions gain, $GVD_{gain.}$ While $GVD_{mat}$ and $GVD_{mod}$ provide a static contribution to the dispersion, intersubband transitions give rise to a bias-dependent dispersion arising from the related change of the gain profile.

Heterogeneous active region (AR) designs provide an effective mechanism to achieve a low and flat $GVD_{gain}$ at the centre of the gain curve [9,14]. Unlike typical homogeneous AR designs, the performance of which are characterized by a narrow gain, modulated by a Lorentzian function, in a heterogeneous AR, the gain is indeed the result of the weighted sum of the gain profiles of each individual active region module. This provides the low and flat intrinsic dispersion required to preserve comb formation in THz QCLs [18,19]. Through use of a metal-metal waveguide, an almost homogeneous field distribution is achieved across the active region, as well as constant waveguide losses and group velocity dispersion over a broad frequency range (1–5 THz), which furthermore assists stable frequency comb operation over a large band.

However, despite these benefits, metal-metal waveguides suffer from the presence of high-order lateral modes that are not completely suppressed and lead to an uneven distribution of power in the fundamental cavity modes over the QCL gain bandwidth [20]. These modes can also be responsible for sub-pulse formation, which is undesirable for frequency comb operation and for mode-locking.

An elegant approach to circumvent this problem is to add side absorbers along the waveguide edges, which inhibits lasing of the higher order lateral modes by increasing their threshold gain [21]. This can be achieved by slightly reducing the width of the top metal layer with respect to the laser ridge, and by covering the underlying GaAs surface with a very thin lossy metallic layer [21]. Using such an approach, broadband QCLs showing frequency comb operation [22] over a limited current range ≈10–30 mA and with a current density dynamic range $J_{dr} = J_{max}/J_{th} \leq 2$ have been demonstrated [9,22], where $J_{th}$ is the threshold current density and $J_{max}$ is the maximum working current density.

In this work, we engineer a broadband heterogeneous QCL to achieve $J_{dr}$ = 3.2, and demonstrate THz QCL frequency comb operation in continuous-wave with an optical power of 4 mW. The beatnote map unveils a clear frequency comb regime over a continuous current range of 106 mA, significantly larger than the state-of-the-art (10-30mA) [9,22], with a current tuning coefficient ranging between 5.0 MHz/mA and 2.1 MHz/mA, and a temperature-tuning coefficient of –4 MHz/K.

The design of the three individual ARs forming the heterogeneous gain medium is a rescaled version of the AR design described in Ref. 23. The number of periods, the order of the AR modules, and the doping $n_d$ were carefully arranged to give a flat gain and uniform power output across the whole spectrum. The final gain medium includes in sequence: 40 periods of a 3.5 THz design, 40 periods of a 3.0 THz design, and 55 periods of a 2.5 THz design. The doping $n_d$ was set to $3 \times 10^{16}$ cm$^{-3}$. Further delays can be found in Ref. 10.

The heterogeneous GaAs/AlGaAs heterostructure used in this work comprises three active modules, grown on a semi-insulating GaAs substrate by molecular beam epitaxy, exploiting alternating photon- and longitudinal optical (LO) phonon-assisted transitions between inter-miniband [25]. Although heterogeneous AR are widely used to achieve broadband frequency combs, broadband THz emission is normally restricted to a small fraction of $J_{dr}$ owing to the differences in $J_{th}$ and $J_{max}$ between individual ARs. Full broadband operation can only be achieved when the injected current density is higher than the $J_{th}$ of each individual AR. In our GaAs/AlGaAs heterostructures, individual ARs possess comparable $J_{dr}$ to ensure that broadband THz emission extends to the highest possible current density dynamic range.

To understand how the side absorbers affect the threshold gain of the fundamental mode, we calculated the frequency-dependent threshold gain of the $TM_{00}$ and the $TM_{01}$ optical modes [21] in metal-metal waveguides of different ridge widths and with different side-absorber dimensions using a finite element method (Comsol Multiphysics). Figure 1a shows the simulated threshold gain $g_{th}$ of the $TM_{00}$ and $TM_{01}$ optical modes over the heterogeneous QCL gain bandwidth for a laser bar of 85 μm width assuming that the $TM_{00}$ and $TM_{01}$ optical modes experience the same mirror losses. Owing to the high confinement provided by the MM waveguide, the threshold is approximately the same for the $TM_{00}$ and $TM_{01}$ modes in the absence of the side absorbers (solid lines), confirming that both modes will reach lasing threshold. By implementing side metal absorbers (dashed lines), the higher order lateral modes experience larger losses than the $TM_{00}$ mode. The thinner absorber (5 μm) leads to a lower $g_{th}$ for the fundamental mode than the 7-μm-wide absorber. We then performed additional simulations to determine the optimum absorber width, for different ridge widths, needed to ensure that the threshold gain of the $TM_{00}$ mode is ~1.5 times smaller than that of the $TM_{01}$ mode. Our simulations show (Fig. 1b) that for laser bars 40–90 μm wide, 2–5-μm-wide absorbers are needed, in agreement with previous reports [21].

Sample fabrication was based on a standard metal–metal processing technique that relies on Au-Au thermo-compression wafer bonding of the 17-μm-thick active region onto a highly doped GaAs substrate. Laser bars were defined using dry-etching, leading to almost vertical sidewalls (and hence uniform current injection into the full gain region). Dry-etching technique also allows the realization of narrow laser cavities without unpumped regions, favoring operation in CW and the creation of a large number of fundamental Fabry– Pèrot modes. Following etching, a Cr/Au (8 nm/150 nm) top contact was deposited along the center of the ridge surface, leaving a thin region uncovered along the ridge edges. 3–5-μm-wide Ni (5-nm-thick) side absorbers (depending on the ridge width) were then deposited over the uncovered region using a combination of optical lithography and thermal evaporation. Laser bars 60–85 μm wide and 2.9 mm long were finally cleaved and mounted on a copper bar on the cold finger of a helium cryostat. A scanning electron microscope image of a prototypical device is shown in Figure 1c.

Figure 2 shows the voltage-current density (V-J) and the light-current density (L-J) characteristics acquired while driving a 2.9 mm long × 85 μm wide QCL in pulsed mode (200 ns pulse width at 100 kHz repetition rate, with an additional 33 Hz electrical envelope modulation to enable lock-in detection of the output power). The QCL has a threshold current density $J_{th}$ = 150 $A/cm^2$, and a maximum peak optical power of 17 mW and exhibits laser action up to 120 K.

Continuous wave LJV characteristics (Fig. 3a) show a maximum power of 4 mW at 20 K with a comparable $J_{th}$.

Spectral measurements were performed using a Fourier- transform infrared (FTIR) spectrometer operating under vacuum, with a resolution of 0.075 $cm^{-1}$. The lasing spectra broaden gradually at increasing the bias current, starting with a ~1.05 THz bandwidth with a few modes missing on the low energy side of the spectrum (Fig.3b), at a current density 100A/$cm^2$ above threshold, and reaching the broadest bandwidth of 1.3 THz (2.25-3.55 THz) at a current density of 450 A/$cm^2$ at 15 K in both pulsed (Fig. 2b) and CW (Fig. 3c) regimes. The spectra present an almost uniform power distribution between all modes within the spectral bandwidth, indicating that the side absorbers effectively reduce any mode competition effects.

To characterize the spectral emission and its coherence, we performed beatnote measurements at different points on the L-I curve. The QCL was driven in CW by a low-noise power supply (ppqSense) and the radio frequency signal was recorded using a bias-tee connecting the QCL and a radio frequency spectrum analyzer (Rohde & Schwarz). The beatnote map is shown in Figure 4a. A prototypical beatnote spectrum at $I$ = 530 mA (215 A/$cm^2$) is shown in Figure 4b. The retrieved narrow beatnote linewidth (4.15 kHz) suggests that the FWM process is locking the lasing modes which are therefore evenly spaced by the cavity roundtrip frequency $f_r$ = c/2Ln, being L the cavity length and n the effective refractive index.

By investigating the evolution of the spectra (Figs 3b-c) and the corresponding RF signals (Fig. 4a) across the THz QCL dynamic range, we can unveil the intracavity dynamic behavior of the laser optical modes. Remarkably, for currents just above threshold (Fig. 3a) and for a continuous 106 mA current range (430mA-536mA), a single narrow beatnote (width < 5KHz) is observed (Fig. 4b and inset of Fig.5a). The corresponding QCL FC spectral bandwidth cover a discontinuous 1.05 THz range and a continuous 0.7 THz range in which 55 equally spaced optical modes regularly distribute. It is worth mentioning that the QCL linewidth is limited here by the jittering of the beat-note because the laser is not stabilized. At higher currents, and for a very limited current range (536–540 mA), a transition to a multi-beatnote regime with an overall mode spacing of 200 MHz can be observed. (Fig.4a). This is likely to be due to some higher order lateral modes reaching threshold in this regime, despite the side absorbers, introducing a slightly different $f_{rep}$ with respect to the fundamental modes. In this condition, the beatnote turns again single and narrow (4.1 kHz), indicating the occurrence of a new comb regime (540–543 mA), prior to entering a dual beatnote regime (544–560 mA) with two dominant beatnotes (14.1 GHz and 14.9 GHz) both with a ~ 4.2 kHz linewidth.

We tentatively attribute this behavior to the dual comb nature of our heterogeneous active core. Indeed, the spectrum in Fig. 3b shows two main families of modes, one centered at ~ 3.1 THz and a second one centered at ~ 2.7 THz that arises at slightly higher applied voltages and whose intensity progressively increases with bias. The two families of modes arise from two of the three active region modules interleaved together [10], which by themselves, behave like a comb over a small current range above their slightly different threshold currents [10].

At larger currents (> 560 mA) we observe a broad beatnote that progressively broadens further upon increasing the applied voltage (Fig. 4a); this feature is characteristic of a lasing regime where the GVD is large enough to prevent the FWM from locking the lasing modes in frequency and phase simultaneously. The large dispersion is here due both to $GVD_{mat}$ and the $GVD_{gain}$. Indeed the heterogeneous nature of the active region entangles the dispersion dynamics at large biases. The corresponding spectrum at $I$ = 1.1 A (Fig. 3c) extends over a 1.3 THz bandwidth. Such a complex mode dynamic is well-known [14,9,22]. While the gain is clamped to the total losses immediately above threshold, the line shape of the gain curve of the heterogeneous active region experiences changes with the applied electric field, meaning that the dispersion is bias dependent.

To further corroborate our data, we measured the electrical and thermal tuning of the beatnote in the operational regime in which the QCL behaves like a frequency comb, during CW operation. Figure 5a shows the tuning of the RF-signal of the THz QCL comb as a function of the driving current and at a fixed operating temperature of 15 K. The beatnote blue shifts across the explored region, spanning a 370 MHz range with a tuning coefficient of 5.0 MHz/mA, which decreases to 2.1 MHz/mA when the current is larger than 480 mA. The observed dependence of the intermode beatnote frequency on pump current cannot be explained by considering only the ohmic heating of the device core, which would conversely predict a monotonic decrease in frequency. The observed trend is likely related to the gain dynamics behind the active region architecture. Specifically, this is a consequence of the variation of the relative distance between the beating modes owing to the variation in effective refractive index, which affects the gain spectrum.

In semiconductor lasers, the refractive index variations (and corresponding emission frequency variations) due to the gain change with pump current can be estimated via the linewidth enhancement factor that unlike diode lasers, in a QCL can be either positive or negative, depending on the position of the emission frequencies relative to the gain peak and the dynamics of the gain spectrum. Finally, it is worth mentioning that the beatnote linewidth remains practically unperturbed over this current range (Inset to Fig. 5a).

Figure 5b shows the dependence of the electrical beatnote on heat sink temperature, while keeping the QCL pump current at 450 mA. The frequency decreases monotonically with an average slope of approximately –4 MHz/K and spans about 98 MHz between 10 K and 34 K. This is a clear effect of the ohmic heating of the laser cavity.

In conclusion, we have developed an ultra-broadband THz QCL that operates as a frequency comb without requiring any additional mode-locking. The beatnote map reveals stable comb behaviour over a 106 mA current range, with a sharp RF signal having an average linewidth of ~ 4.2 kHz. The emission bandwidth spans 1.3 THz in both continuous wave and pulsed mode operation. The demonstrated current and temperature-induced comb tuning that has been demonstrated provides useful insight on the suitability of the THz QCL combs as metrological tools for high-resolution spectroscopy at THz frequencies and as local oscillators for heterodyne detection.

**Acknowledgements**

We acknowledge financial support from the EC Project 665158 (ULTRAQCL), and the ERC Project 681379 (SPRINT), the EPSRC Programme Grant EP/P021859/1 (HyperTerahertz) and the Royal Society project IEC/NSFC/170384. EHL acknowledges the support of the Royal Society and Wolfson Foundation.

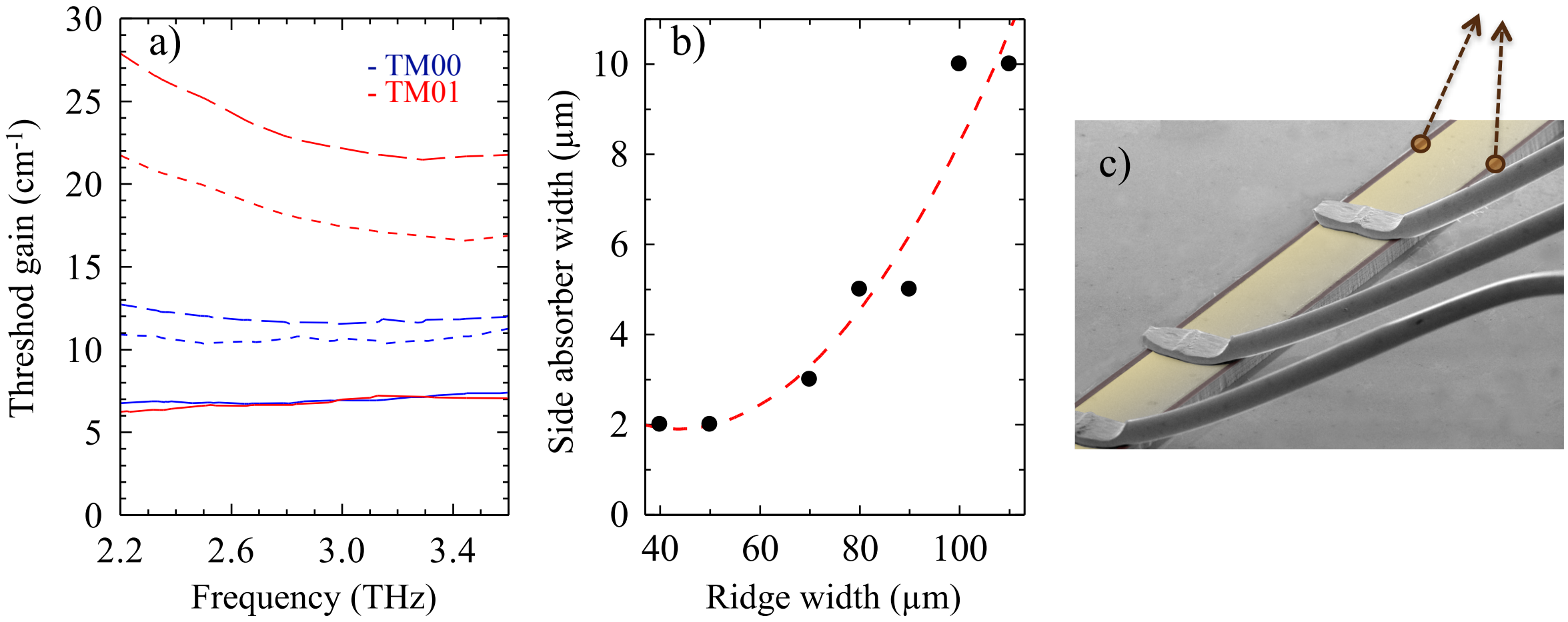

**Figure 1.** a) Simulated threshold gain as a function of frequency for a 85-μm-wide laser ridge embedded in a metal-metal waveguide. The impact of the lossy side absorbers is evaluated by comparison between the threshold gain of the $TM_{00}$ (blue lines) and $TM_{01}$ (red lines) optical modes when no (solid lines), 5-μm-wide (dotted lines) and 7-μm-wide (dashed lines) nickel side absorbers are implemented on both sides of the ridge. b) Calculated optimal side absorber width as a function of ridge width. The optimal side absorber width is evaluated by calculating the smallest width that would give a threshold gain for the $TM_{01}$ mode that is ~1.5 times smaller than that of the $TM_{00}$ mode. The dashed line is a fit to the data. c) False colour scanning electron microscope image of a fabricated laser bar. The Ni side absorbers are indicated.

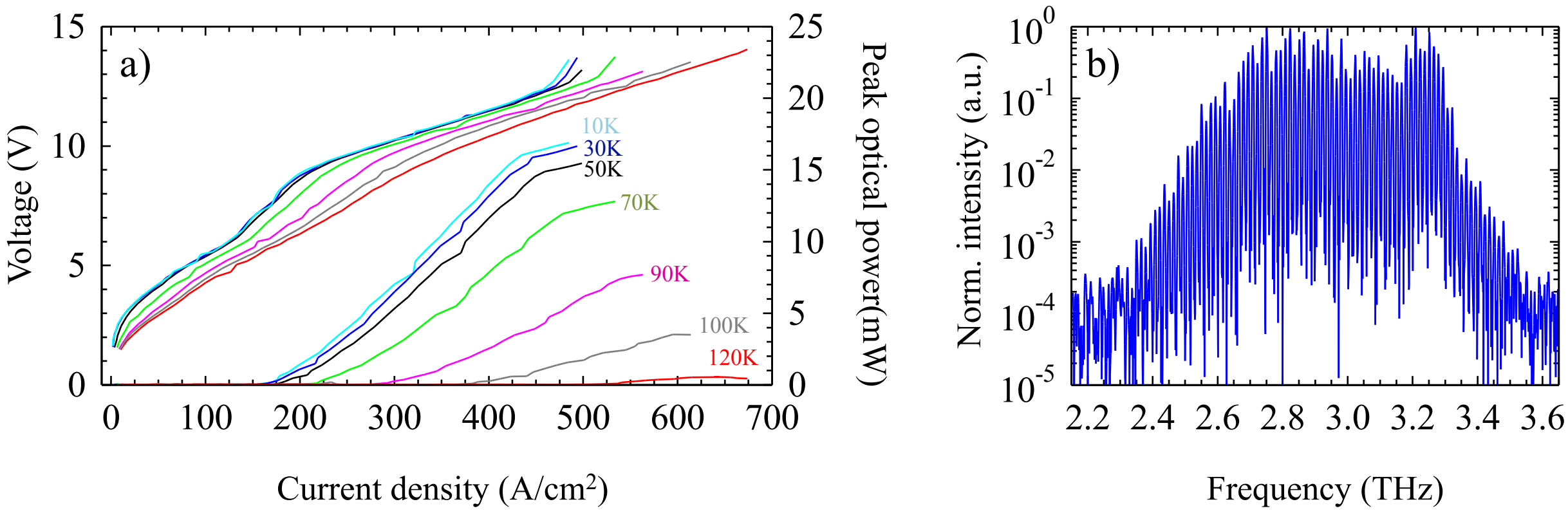

**Figure 2. a)** Voltage – current density (V-J) and light – current density (L-J) characteristics measured in pulsed mode as a function of heat sink temperature, in a 2.9-mm-long, 85-μm-wide laser bar, with 5-μm-wide side absorbers present on both sides of the laser ridge. A pyroelectric sensor having sensitive area of $7mm^2$ has been employed for measuring the optical power. b) A Fourier transform infrared (FTIR) spectrum collected in rapid scan mode, under vacuum, with a 0.075 cm$^{-1}$ resolution at 15 K, while driving the QCL in pulsed mode under the same experimental conditions as (a).

Figure 3

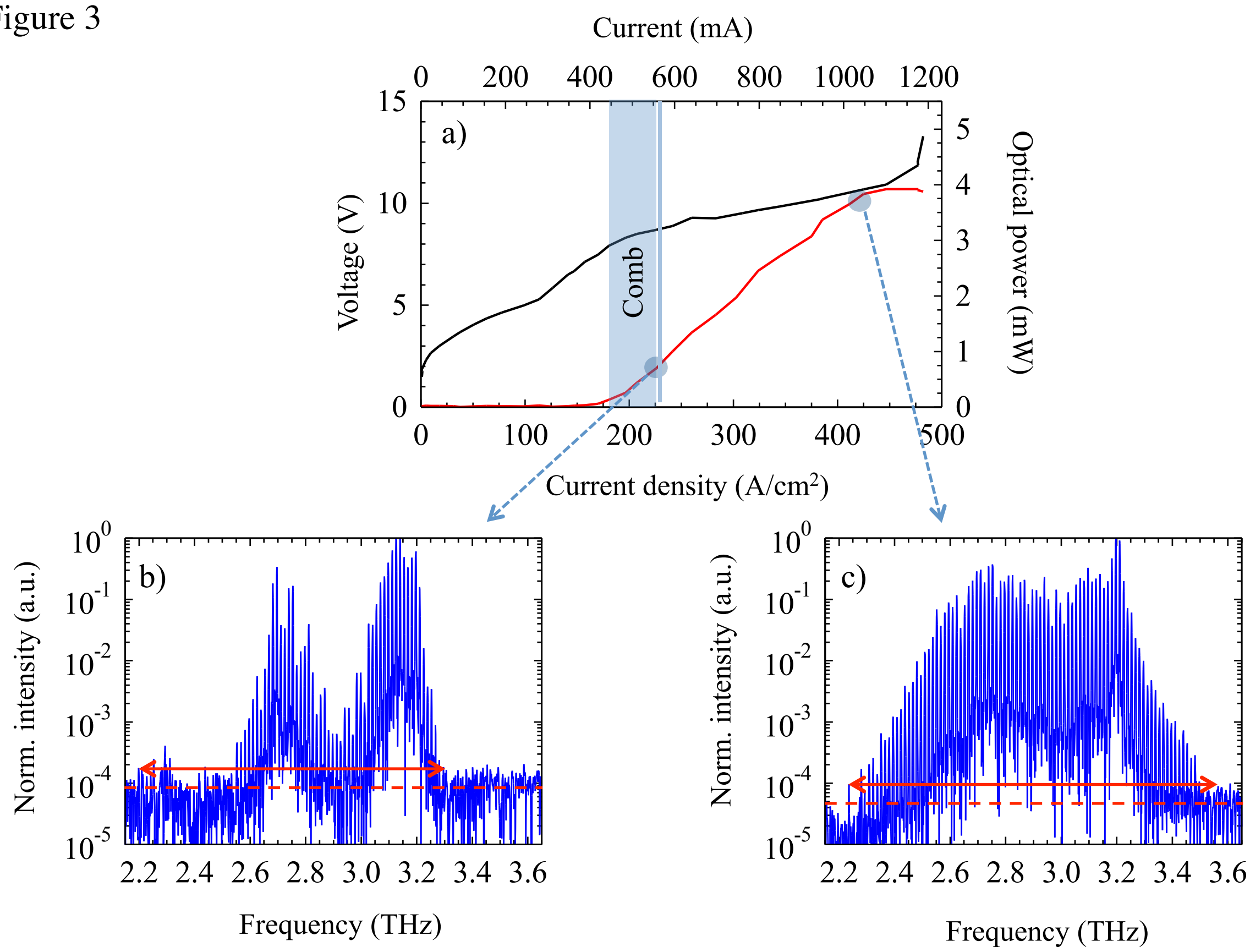

**Figure 3. a)** Voltage - current density (V-J) and light – current density (L-J) characteristics measured in continuous wave, at 15 K in a 2.9-mm-long, 85-µm-wide laser bar, with 5-µm-wide side absorbers present on both sides of the laser ridge. The two light blue rectangular areas indicate the operational regimes of the frequency comb. The optical power has been measured using a broad-area terahertz absolute power meter (TK Instruments, aperture 55 × 40 mm$^2$). b-c) FTIR spectra collected in rapid scan mode, under vacuum with a 0.075 cm$^{-1}$ resolution at 15 K, while driving the QCL in continuous wave at (b) 536 mA and (c) 1.1 A, corresponding to the light blue dots on the J-V characteristic of panel (a). The dashed horizontal lines roughly indicate the noise floor of the measurements. The red arrows mark the emission bandwidth.

Figure 4

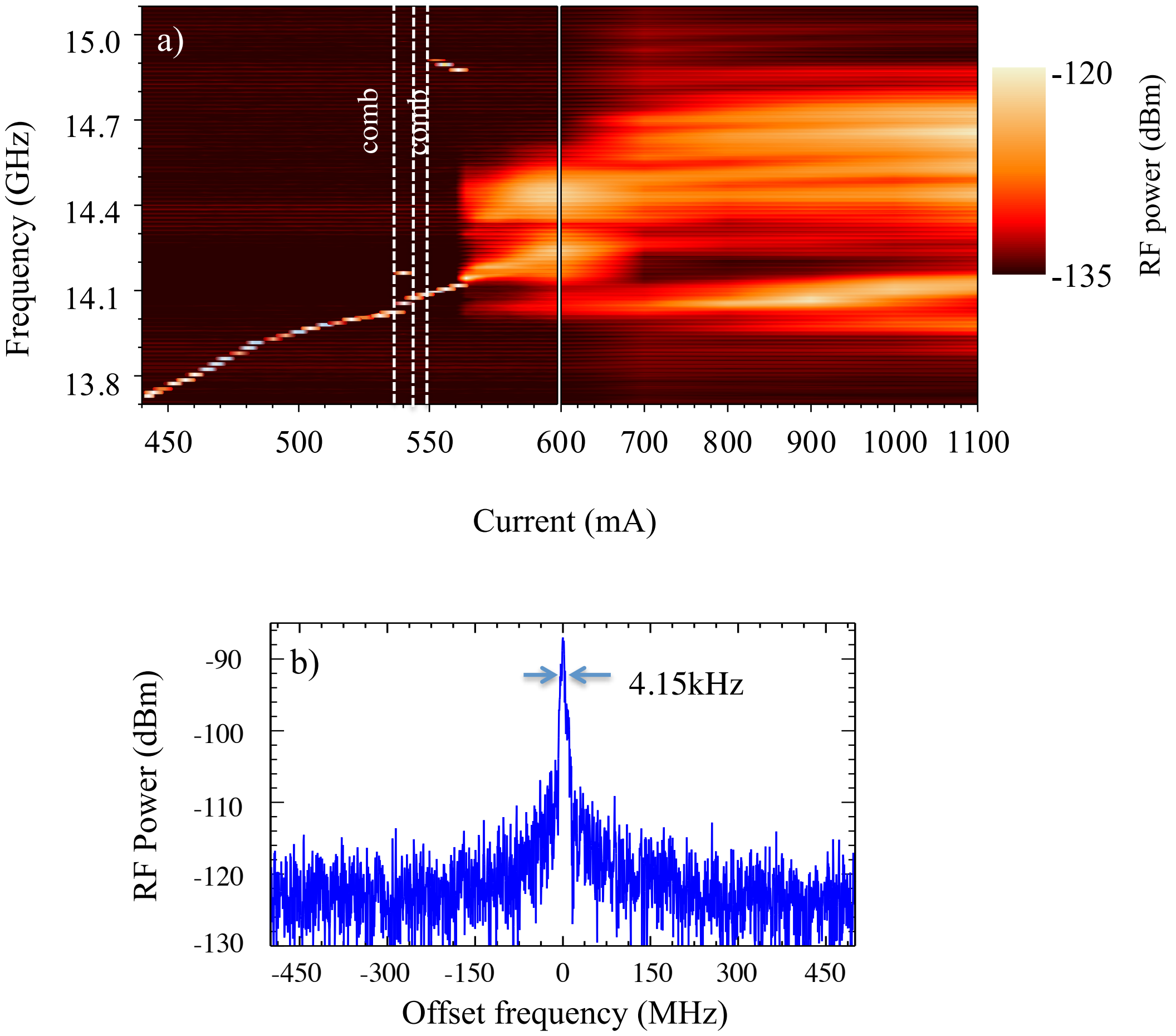

**Figure 4. a)** Intermode beatnote map as a function of driving current obtained at 15 K from a 2.9-mm-long, 85-µm-wide laser bar. The beatnote signal is extracted from the bias line with a bias-tee and is recorded with an RF spectrum analyser (Rohde & Schwarz FSW; RBW: 500 Hz, video bandwidth (VBW): 500 Hz, sweep time (SWT): 20 ms, RMS acquisition mode). The dispersion compensated comb regime is marked on the graph; dashed vertical lines identify the ranges in which the device behaves as a single frequency comb; b) Intermode beatnote of the same laser under the same experimental conditions of panel (a) recorded with an RF spectrum analyzer (RBW: 10 kHz, VBW: 100 kHz, SWT: 500 ms). The linewidth is limited by the RBW of the spectrum analyzer.

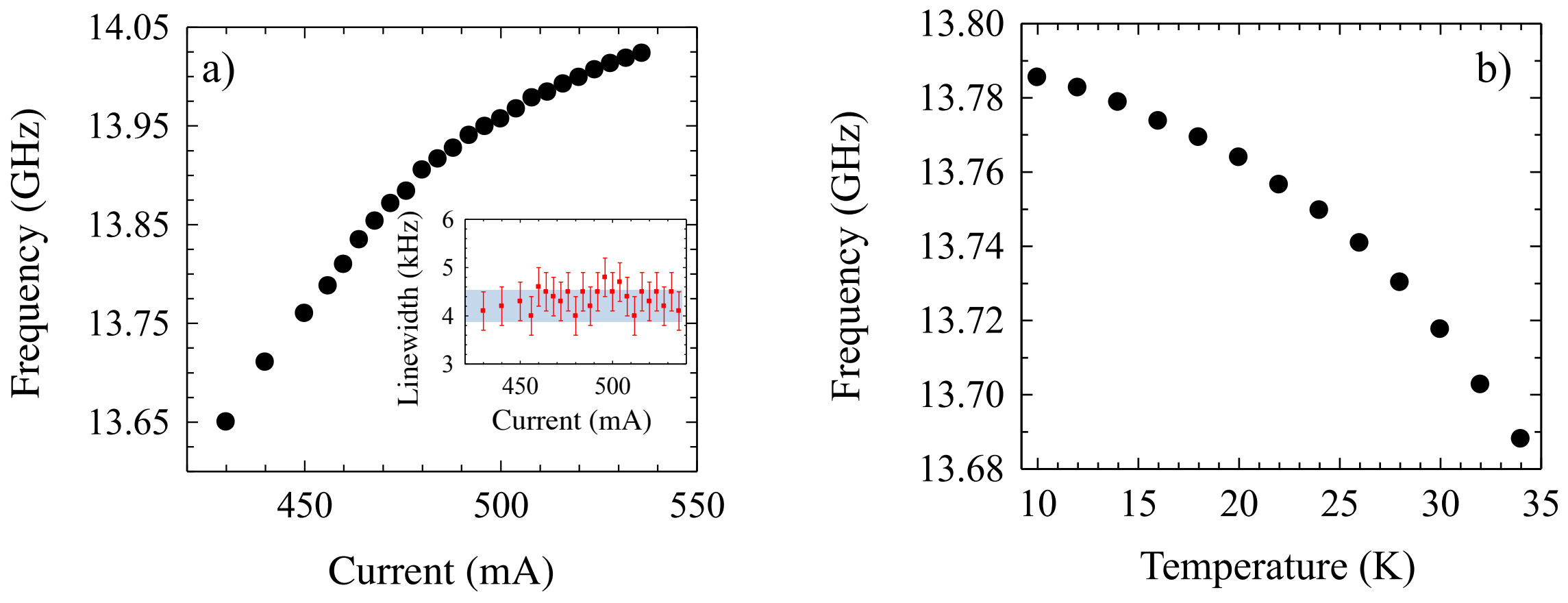

**Figure 5. a)** Intermode beatnote frequency as a function of the continuous wave driving current measured at 15K. Inset: corresponding linewidth of the intermode beatnote frequency. b) Intermode beatnote frequency as a function of the temperature, measured at 15 K while driving the QCL in continuous wave at a current of 450mA.